\documentstyle[prb,aps,multicol,epsfig]{revtex}

\begin{document}

%\draft

\title{A critical reexamination of the non-reciprocal x-ray gyrotropy in 
V$_2$O$_3$. }

\author{S. Di Matteo$^{a,b}$, Y. Joly$^{c}$, C.R. Natoli$^{a}$}
\address {$^a$INFN Laboratori Nazionali di Frascati, c.p. 13, I-00044 
Frascati, Italy \\
$^b$INFM Udr Roma III, via della Vasca Navale 142, 00100 Roma, Italy \\
$^c$Laboratoire de Cristallographie, CNRS, B.P. 166, F-38042, Grenoble
Cedex 9, France}

\date{\today}

\maketitle

\begin{abstract}

We critically review the experimental determination of the nonreciprocal x-ray gyrotropy in the monoclinic antiferromagnetic insulating phase of V$_2$O$_3$ highlighting some of its paradoxical consequences. We also note the apparent absence of a normal (non magnetic) linear dichroism  in the data and speculate on the possible reasons for it. Given the importance of the implications of a non reciprocal effect for the physics of V$_2$O$_3$, one of the paradigmatic strongly correlated system, we point up the need for at least repeating the experiment under better controlled conditions and suggest possible tests for establishing unambiguously its existence.

\vspace{0.3cm}

%\noindent Keywords: x-ray scattering, magnetic space groups, Mott system

\end{abstract}

%\nopagebreak

\begin{multicols}{2}

%text

\section{Introduction}

The crystal and electronic structure of V$_2$O$_3$ has been the subject 
of an intensive experimental and theoretical study throughout
the 70's \cite{mcwhan1,dernier,moon,weger,cnr1,cnr2} that led to the 
identification of V$_2$O$_3$ as the prototype of Mott-Hubbard systems. This  
compound shows a metal-insulator transitions, due to the interplay between
band formation and electron Coulomb correlation,
from a paramagnetic metallic (PM) phase at room temperature to a paramagnetic 
insulating (PI)
phase at higher temperature ($\approx 500~ K$) and to an antiferromagnetic 
insulating (AFI) phase at lower temperature
($T_c \approx 150~ K$ for the stoichiometric compound). Associated with this
latter there is also a structural transition from a corundum cell
(rhombohedral system) to a monoclinic one (see Fig. \ref{cell}). 

In this investigation a correct description of the properties of the AFI 
ground state has been of paramount importance for both the physical 
understanding 
of the insulating phase and for throwing light on the mechanism of the 
metal insulator transition. Indeed Castellani, Natoli and Ranninger (CNR), 
\cite{cnr1,cnr2} were the first to realize the importance of the orbital 
degrees of freedom for the explanation of the peculiar antiferromagnetic 
spin structure \cite{moon} that breaks the trigonal symmetry of the corundum 
lattice in the high temperature phase. Evidence for their role was subsequently
found in a series of inelastic neutron scattering experiments by Wei Bao 
{\it et al.} \cite{weibao} and NMR nuclear spin relaxation measurements 
\cite{taki} in the high temperature paramagnetic phases. The ensuing physical 
picture was one in which the orbital degrees of freedom are frozen in the AFI 
phase, giving rise to the peculiar spin structure, and are responsible for 
the short range magnetic fluctuations in the high temperature phases, 
providing much of the necessary entropy content for the transition to occur.
Even though the Hubbard model used by CNR was based on a spin S = 1/2 for the 
$V^{3+}$ ion in a doubly degenarate orbital $e_g$ state, the subsequent 
picture based on spin S=1, implied by some experiments 
\cite{paolasini1,park} and worked out by Mila {\it et al.} 
\cite{mila1,mila2} and Di Matteo {\it et al.},\cite{dimatteo} 
did not change the above physical description, but only redefined the basic 
building blocks of the problem. Indeed in  the S=1 picture the elementary 
units are the vertical biatomic molecules in a doubly degenerate ground 
state, strongly coupled to spin S=2, and weakly interacting via electronic 
kinetic hopping in the basal planes. 

The introduction of the extra degrees of freedom in the problem might 
be expected to change the magnetic space group of the ordered AFI phase as 
inferred from the lattice geometry and spin moments only. As explained in 
the following section, the magnetic space group of the monoclinic 
cell with the atomic sites occupied by the magnetic moments in the way found 
by Moon \cite{moon} is the magnetic space group $P2/a+{\hat T}\{{\hat E}|t_0\}P2/a$,\cite{cracknell} containing explicitly the time reversal operator ${\hat{T}}$ 
and the inversion $I$ (${\hat E}$ is the identity operator, $t_0$ is the body-centered translation). 

In this context the observation of a nonreciprocal x-ray linear dichroism 
measured at the K-edge of the Vanadium atom in the AFI phase of V$_2$O$_3$ 
\cite{goulon} seemed to shed new light on the properties 
of the insulating ground state, since the effect can only be observed when 
neither the inversion nor the time reversal are separately symmetry 
operations of the system, thus implying that the 
system should be magnetoelectric. Even though not explicitly stated in Ref. 
[\onlinecite{goulon}] this finding was pointing to a role of the orbital 
degrees of freedom in lowering the magnetic symmetry of V$_2$O$_3$, all the 
more that an independent observation of the forbidden (111)$_m$ 
reflection in the $V$ K-edge anomalous diffraction \cite{paolasini1}
had been interpreted as implying some kind of orbital ordering. 

However a closer consideration of all the pieces of evidence of the problem 
revealed some inconsistencies in the overall interpretative frame, 
as discussed at length in Ref. [\onlinecite{dimatteo}]. First of all 
the theoretical model used in Ref. [\onlinecite{paolasini1}] to interpret  
the orbital ordering was based on the old CNR S=1/2 model, which was later 
shown to be inadequate to describe the physics of V$_2$O$_3$. In the new 
correlated model based on spin S=1 \cite{mila1,mila2,dimatteo} the orbital 
ordering in the AFI ground state was again found compatible with the 
$P2/a+{\hat T}\{E|t_0\}P2/a$ group. In the meantime growing evidence was  
accumulating that the forbidden (111)$_m$ reflection is of pure magnetic 
origin and has nothing to do with orbital ordering. 
\cite{lovesey1,tanaka,lovesey2} Last but not least, a new
direct measurement of the magnetoelectric effect in V$_2$O$_3$ has provided 
a negative result, \cite{jansen} confirming the previous result by Astrov. 
\cite{astrov}

Concomitant to all this, our efforts to understand at a deeper level the 
implications of the nonreciprocal x-ray gyrotropy experiment led us to 
note some paradoxical consequences of this result and to look for alternative 
interpretations. 
It is the purpose of this paper to present the conclusions of 
this investigation and to point up at least the need to repeat the experiment 
under better controlled conditions.

This paper is organized as follows. In section II, we discuss in some detail, 
for the convenience of the reader, the structural and magnetic properties of 
the system in its monoclinic phase and derive the corresponding magnetic 
space group, listing all the magnetoelectric subgroups. 
In section III we derive explicitly the response functions corresponding 
to a linear dichroism experiment both in the dipole-dipole (E1-E1) and the 
dipole-quadrupole (E1-E2) channel. We show that the corresponding signal is
proportional to the average of certain operators in the final scattering 
state of the excited photoelectron, having definite transformation properties 
under inversion and time reversal, and discuss in the light of the possible 
magnetic space groups the conditions under which such a signal is different 
from zero. We find that contrary to what implied in Ref. [\onlinecite{goulon}] 
one expects a significant signal from the E1-E1 channel, which is inversion 
and time reversal even. We further argue that its existence is a 
necessary consequence of the Templeton scattering at the 
(10$\overline{1}$)$_m$ forbidden reflection in the $\sigma-\sigma$ channel 
observed by Paolasini {\it et al.}\cite{paolasini2} in the monoclinic phase of the same compound. \cite{nota1} 
We estimate the dichroic signal in the frame of multiple scattering (MS) theory, which should be adequate to describe transitions to the itinerant conduction 
$p$ states in V$_2$O$_3$. 
Surprisingly enough its shape is quite similar to the ``nonreciprocal'' 
gyrotropy signal reported in Ref. [\onlinecite{goulon}]. 
We speculate on such similarity and give arguments whereby the reversal of 
the signal under magnetic field\cite{goulon} could in fact be due to an 
uncontrolled geometrical effect. In section IV we summarize our conclusions.

\section{The monoclinic structure of V$_2$O$_3$.}

Pure V$_2$O$_3$, as well as the 2.8\% chromium doped sample used in the 
x-ray experiments mentioned in the introduction, is a paramagnet 
at room temperature that crystallize in the trigonal system.
At T$_c \simeq$ 150 K (180 K for the doped compound), it undergoes a 
structural transition to the monoclinic system and an antiferromagnetic 
order sets in.
This doubles the unit cell to four 
formula units, instead of the two formula units of the trigonal cell.
Moreover the transition breaks the trigonal symmetry in such a way that 
one of the three originally equivalent vanadium-vanadium bonds in the 
hexagonal plane of the corundum cell becomes ferromagnetic and longer of about
$4\%$ than the other two, which are antiferromagnetic. Because of this, three 
possible monoclinic twin domains exist, depending on which of the three 
equivalent bonds stretches and becomes ferromagnetic. They are related by a 
rotation of 120$^o$ around the corundum c$_H$-axis. Experimentally one finds 
that V$_2$O$_3$ has the tendency to crystallize prevalently in one of the 
three monodomains, more than in a statistical average of the three and that 
the choice among them is more or less random, depending on the size, shape 
and boundary conditions of the crystal.\cite{luigi}

We give here a detailed description of the monoclinic
cells and of its lattice and magnetic  symmetries, as we shall use them
in the subsequent  analysis of the structure factor.

\begin{figure}
\centerline{\epsfig{file=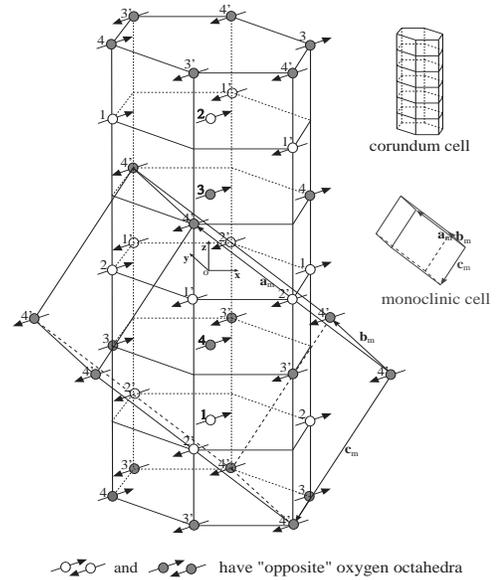,height=3.0in,
width=2.5in}}
\vspace{5pt}
\caption
{Corundum and monoclinic cells for the paramagnetic  
and the antiferromagnetic phases of V$_2$O$_3$, respectively. The $a_H$-vector coincides with $b_m$ and $b_H$ connects 4' with 4.
The distortion and increase in volume due to the monoclinic 
transition are not shown. Note that V-ions do not lay on the hexagonal planes, but they are slightly displaced ($0.17 \AA$) towards the nearest voids.}
\label{cell}
\end{figure}

The crystal positions are taken from Dernier and Marezio,
\cite{dernier}  and the magnetic structure of the monoclinic cell from Moon
\cite{moon} later confirmed by Wei Bao.\cite{weibao}

The monoclinic cell, shown in Fig. 1, belongs to the body-centered crystallographic space group I2/a. 
Using the reference frame and the numbering of vanadium atoms of
Fig. \ref{cell}, we can divide
the eight atoms of the monoclinic cell into two groups of four, with opposite
orientation of the magnetic moment, $V_1 =  (1/2-u, v, -w), \; 
V_2 = (1/2+u, -v, w), \; V_3 = (-u, -v, -w), \; V_4 = (u, v, w) $,
and $V_{1'}$, $V_{2'}$, $V_{3'}$, $V_{4'}$, with coordinates obtained by 
adding the vector (1/2, 1/2, 1/2) to the first group.
Here $u = 0.3438, \; v = 0.0008, \; w = 0.2991$ are the fractional 
coordinates of the atoms in unit of the monoclinic axis.\cite{dernier}
Note that, neglecting the magnetic moments, the two groups of atoms with 
their oxygen environments are translationally equivalent.

From these data we can infer that the magnetic space group of the
monoclinic cell is the magnetic space group $P2/a+{\hat T}\{E|t_0\}P2/a$,\cite{cracknell} where ${\hat {T}}$ is the  
time-reversal operator
and the monoclinic group $P2/a$  contains the identity  ${\hat {E}}$, 
the inversion ${\hat {I}}$, the two-fold  rotation about the monoclinic
b$_m$-axis ${\hat {C}}_{2b}$ and the  reflection ${\hat {m}}_b$ with
respect to the plane perpendicular  to this axis. Of course, the
appropriate translation is associated to  each of these operators.
\cite{dimatteo}
Choosing the origin of the system as in Fig. 1, we have the following 
symmetry operations:

\begin{eqnarray}
\begin{array}{lll}
1){\rm ~} {\hat E}, {\rm ~} {\hat I} &\rightarrow & {\rm No ~translation} \\
2){\rm ~} {\hat C_{2b}}, {\rm ~} {\hat m_b} &\rightarrow &
\frac{1}{2}(\vec{b}_m+\vec{c}_m) \\
3){\rm ~} {\hat T}, {\rm ~} {\hat T \hat I} &\rightarrow &
\frac{1}{2}(\vec{a}_m+\vec{b}_m+\vec{c}_m) \\
4){\rm ~} {\hat T \hat  C_{2b}}, {\rm ~} {\hat T \hat m_b } &
\rightarrow & \frac{1}{2}\vec{a}_m ~.
\end{array}
\nonumber
\end{eqnarray}

Notice that, if one considers only the lattice and magnetic moments, 
the presence of the time-reversal symmetry is a necessary consequence of 
the translational equivalence of the two groups of atoms. 
Now, as anticipated in the introduction, one of the consequences of the 
observation of non-reciprocal effects by linear dichroism\cite{goulon} is 
the indication that other degrees of freedom may play a role in determining 
the symmetry of the insulating ground state. In fact a necessary condition 
to detect the non-reciprocal gyrotropy tensor is that the system be 
magnetoelectric, this fact implying in turn the breakdown of time-reversal 
and inversion symmetry. Orbital degrees of freedom are the natural candidates 
in the physics of V$_2$O$_3$ to obtain such a symmetry reduction, although 
new, still unsuspected, ingredients might come into play.
One other possibility, already debated in the literature,\cite{moon,word,yethirai} yet very controversial, concerns the presence of an out-of-plane (a$_m$-c$_m$ plane) component of the magnetic moment directed along the $y$ axis.. This fact would at least break the glide-plane symmetry.  Three independent neutron scattering experiments\cite{moon,word,yethirai} were not able either to discard or accept this hypotesis. However this out of plane component would contradict the argument given in Refs. [\onlinecite{dimatteo,lovesey1,tanaka,lovesey2}] 
to explain the absence of any dipole signal at the (111)$_m$ reflection in the energy region of the conduction band of $p$ symmetry, that can only be explained by the presence of the glide-plane symmetry together with the fact that $<L_y>=0$. For this reason, we can conclude that x-ray measurements point
toward the absence of an out-of-plane component of the magnetic moment.

Here we list for completeness all the possible space magnetic subgroups of
$P2/a+{\hat T}\{E|t_0\}P2/a$ that do admit magnetoelectricity:\\
\begin{eqnarray}
\begin{array}{lll}
1) {\rm ~} ({\hat E}, {\hat T \hat I}, {\hat T \hat  C_{2b}}, {\hat m_b}) 
    & \equiv  P2'/a & \\
2) {\rm ~} ({\hat E}, {\hat T \hat I}, {\hat  C_{2b}}, {\hat T  \hat m_b}) 
    & \equiv  P2/a' & \\
3) {\rm ~} ({\hat E}, {\hat T \hat I}) & & \\
4) {\rm ~} ({\hat E}, {\hat T \hat C_{2b}}) & &  \\
5) {\rm ~} ({\hat E}, {\hat C_{2b}})  & & \\
6) {\rm ~} ({\hat E}, {\hat T \hat m_b}) & &  \\
7) {\rm ~} ({\hat E}, {\hat m_b}) & & \\
8) {\rm ~} ({\hat E}) & & 
\end{array}
\nonumber
\end{eqnarray}
We have indicated in parenthesis the symmetry operations of the associated 
point groups and for simplicity we have not written the corresponding 
translations which are the same as above. Note that all the two elements 
subgroups actually belong to the triclinic system.
On the basis of these groups we shall discuss in the next section the 
conditions for the observation of linear dichroism in absorption.

\section{Normal and nonreciprocal linear dichroism}

Linear dichroism in the x-ray range is the differential spectrum obtained 
by subtracting two core absorption spectra taken with orthogonal linear 
polarizations. According to whether it is of dipole-dipole (E1-E1) or 
dipole-quadrupole (E1-E2) origin we shall distinguish, for convenience and 
reasons to become apparent later, between respectively:
Normal (NXLD) and Nonreciprocal (NRXLD) X-ray Linear Dichroism. The two 
contributions can obviously be both present in the same spectrum.
The experiment\cite{goulon} we are interested in was performed with 
the x-ray beam directed along the hexagonal c$_h$-axis, or 
(20$\overline{2}$)$_m$ in monoclinic Miller indices, with the two orthogonal 
directions of the linear polarizations lying in the hexagonal plane 
without specification of their orientation with respect to the in plane 
crystallographc axes.

\subsection{Dichroic operators and related sum rules}

As is well known, at the Vanadium K-edge the x-ray absorption cross section 
is given by:

\begin{equation}
\sigma_i=4\pi^2 \alpha \hbar \omega \sum_n |\langle \Psi_n ^{(i)}| 
{ \hat {O}} | \Psi_0^{(i)} \rangle |^2 \delta(\hbar\omega-(E_n-E_0))
\label{absor}
\end{equation}

The operator ${\hat {O}}\equiv {\hat {\epsilon}} \cdot \vec{r} 
(1 + \frac{i}{2} \vec{k} \cdot \vec{r})$ is the usual matter-radiation interaction 
operator expanded up to the quadrupolar term, with the usual notation for the 
photon polarisation ${\hat {\epsilon}}$ and the wave vector 
$\vec{k}$. $\Psi_0^{(i)}$ ($\Psi_n^{(i)}$) is the ground (excited) state of 
the crystal, $E_0$ ($E_n$) its energy and the index $i$ indicates the lattice 
site of the Vanadium photoabsorbing atom. 
The sum is extended over all the excited states of the system and $\hbar\omega$ 
is the energy of the incoming photon, while $\alpha$ is the fine-structure 
constant.

Before analyzing Eq. (\ref{absor}) on the basis of the crystal space magnetic symmetry, we derive explicitly the cross section and the consequent sum rule for non-reciprocal and normal linear dichroism, on the basis of the formalism developed in Ref. [\onlinecite{dimatoli}].
In both cases linear dichroism is defined as: 

\begin{equation}
\sigma_i^{LD} \equiv \sigma_i (\epsilon_a) - \sigma_i (\epsilon_b)
\label{lindicr}
\end{equation}

\noindent where $\epsilon_a$ and $\epsilon_b$ are ususally chosen orthogonal. 
  With reference to Sec. 7 of Ref. [\onlinecite{dimatoli}], 
we can write the signal up to the E1-E2 channel as the scalar product of two tensors, representing respectively the properties of the light (T) and of the matter (M):

\begin{eqnarray}
\sigma_{i}^{LD} \propto \sum_{p,q} (-)^{1+p+q} [ T^{(p)}_q (\epsilon_a, 
\epsilon_a, k) M^{(p)}_{-q} \nonumber \\ 
 - T^{(p)}_q (\epsilon_b, \epsilon_b, k) M^{(p)}_{-q}]
\label{magchir1}
\end{eqnarray}

\noindent where p=1,2,3. 
The explicit expression of the tensors $T^{(p)}_q$ is given in  Sec. 7 of Ref. [\onlinecite{dimatoli}]. Moreover, in terms of the tensors expressing the properties of the matter (M, for E1-E1 and ${\hat M}$ for E1-E2) we obtain:

\begin{equation}
\sigma_i^{NXLD} \propto \frac{1}{\sqrt{5}} (M^{(2)}_2 + M^{(2)}_{-2})
\label{lindicr2}
\end{equation}

\begin{eqnarray}
\sigma_i^{NRXLD} \propto (({\hat M}^{(3)}_2 + {\hat M}^{(3)}_{-2} - c.c. ) + \nonumber \\ 
\frac{1}{\sqrt{2}} ({\hat M}^{(2)}_2 - {\hat M}^{(2)}_{-2} - c.c.))
\label{lindicr3}
\end{eqnarray}

In both sets of equations we have omitted for simplicity the index $i$ in the right hand side.
To evaluate the matter tensor we shall adopt the one particle approach in the framework of multiple scattering theory. However, as shown in Ref. [\onlinecite{dimatoli}], all the expressions remain valid in a second quantization scheme, if one takes the multiple scattering functions as the one particle basis for the matter operators. Around each site the photoelectron wave function can be expanded in spherical harmonics and spin states as: 
\begin{equation}
|\Psi_i({\vec{k}_e})> = \sum_{l,m,\sigma} B^i_{lm\sigma}(\vec{k}_e)R^i_{lm\sigma}(r;E) Y_{lm}(\hat{r})\chi_\sigma 
 \label{psi}
\end{equation}

\noindent where $\vec{k}_e$ is the photoelectron wave vector associated to the final scattering state with kinetic energy $E\propto k_e^2$ and $\chi_\sigma$ is the spin state. Using this expression it is possible to write the tensors $ M^{(p)}_{q}$  in terms of the scattering amplitudes $ B^i_{lm\sigma}(\vec{k})$. In the case of interest (K-edge) $l$ is either 1 or 2 so that we find in the geometrical setting of the 
experiment of Ref. [\onlinecite{goulon}]:

\begin{eqnarray}
\sigma_i^{NXLD}& =&4\pi^2 \alpha \hbar \omega M^2_l(E) \int d\hat{k}_e 
\sum_{m,\sigma} |m|  (B^i_{1m\sigma})^* B^i_{1,-m\sigma} \nonumber \\ 
 & & \propto \int d\hat{k}_e <\Psi_i({\vec{k}_e})|(L^2_{x'}-L^2_{y'})|\Psi_i({\vec{k}_e})>
\label{lindice1e1}
\end{eqnarray}
\begin{eqnarray}
 &\sigma_i^{NRXLD} =4\pi^2 \alpha \hbar \omega M^2_l(E) \int d\hat{k}_e  \nonumber \\ 
      &   \sum_{m,\sigma} |m| (B^i_{2m\sigma})^* B^i_{1,-m\sigma}  - c.c.               \nonumber \\  
      &   \propto  \int d\hat{k}_e 
<\Psi_i({\vec{k}_e})|(L^2_{x'}-L^2_{y'})\Omega_z|\Psi_i({\vec{k}_e})> - c.c.
\label{lindice1e2}
\end{eqnarray}

\noindent where, indicating by $R_{mt}$ the muffin-tin radius, the proportionality factor is 

$$\bigg[\frac{M_l(E)}{\int_0^{R_{mt}}r^2 dr R^2_l(r;E)}\bigg]^2~.$$

In the previous equations we have ignored the weak $m\sigma$ dependence of the radial wave functions, as justified for deep core (1s) electron transitions. We have introduced the dipole radial transition matrix element, $M_l(E)$ and integrated over all the direction of the escaping photoelectron (${\hat{k}}_e$) at a fixed photon energy, as appropriate for absorption. \cite{dimatoli}  
Moreover we have used the equality $(L^2_x-L^2_y)= (L^2_{+} + L^2_{-})$ and the fact that the magnetic quantum number $m$ runs from --1 to 1. The labels $x'$ and $y'$ indicate two arbitrary orthogonal directions of the two polarizations $\epsilon_a$ and $\epsilon_b$ in the hexagonal plane.
As discussed in Ref. [\onlinecite{dimatoli}] the energy integrated spectrum provides the expectation value of the same operators over the ground state, a situation that is in common with all kinds of dichroisms.

Equations (\ref{lindice1e1}) and (\ref{lindice1e2}) show
that linear dichroism at the energy $\hbar \omega$ of the incoming photon measures the expectation value of a physical operator over the final scattering state corresponding to that energy. In particular, this operator is proportional to $L^2_{x'}-L^2_{y'}$ in the dipole-dipole channel and to $(L^2_{x'}-L^2_{y'}) \Omega_{z}-c.c.$ in the dipole-quadrupole channel.
In the last expression $\Omega_{z}\equiv ({\vec {L}} \times {\hat {n}}-{\hat {n}} \times {\vec {L}})_{z}$, where ${\vec {L}}$ is the usual angular momentum operator and ${\hat {n}}$ the radial unit vector operator. 
It is the raising/lowering operator for the orbital quantum number $l$ when applied to a spherical harmonic with the same $m$ projection, \cite{varshalo} and is known as the anapole\cite{carra}  or toroidal moment operator. 

Note that the linear dichroism signal is time-reversal even and inversion even in the dipole-dipole channel, while it is time-reversal odd and inversion odd in the dipole-quadrupole channel, as expected on the basis of general considerations. 

\subsection{Extinction rules for linear dichroism.}

In the monoclinic crystal, the total absorption signal is obtained by summing over all the eight inequivalent  vanadium sites $I_C =  \sum_{i=1}^8  \sigma_i$ in  the unit cell. Therefore, indicating with the label $LD$ both types of dichroism we obtain 

\begin{eqnarray}
I_{LD} = \sum_{i=1}^8 \sigma_i^{LD}
\label{intensity}
\end{eqnarray}

By exploiting the symmetry operations of the system, we can now relate one another the various $\sigma_i^{LD}$ in the unit cell, exactly as done for the anomalous scattering amplitude in Ref. [\onlinecite{dimatteo}]. Absorption is in fact the imaginary part of the forward scattering amplitude. A symmetry operation then acts both on the site index and on the operator appearing in the averages (\ref{lindice1e1}) and (\ref{lindice1e2}), assuming that the final scattering state $|\Psi_i({\vec{k}_e})>$ does not break the symmetry of the ground state. This latter assumption is physically plausible since in the one electron approximation the rest of the system is seen by the excited photoelectron as a static scattering potential.
Note that the symmetry considerations in this subsection do not depend on the one particle approximation introduced in the previous subsection, but hold a more general validity, relying only on the symmetry group of V$_2$O$_3$..
 
The site transformation rules for the largest magnetic group $P2/a+{\hat T}\{E|t_0\}P2/a$ are given in the following table:

\begin{center}
$\begin{array}{||l|c|c|c|c|c|c|c|c||}  \hline
{\hat E}:          & \sigma_1 & \sigma_2 & \sigma_3 & \sigma_4 & \sigma_{1'} & \sigma_{2'} & \sigma_{3'} & \sigma_{4'}  \\ \hline
{\hat I}:          & \sigma_2 & \sigma_1 & \sigma_4 & \sigma_3 & \sigma_{2'} & \sigma_{1'} & \sigma_{4'} & \sigma_{3'}  \\ \hline
{\hat C}_{2b}:        & \sigma_{4'} & \sigma_{3'} & \sigma_{2'} & \sigma_{1'} & \sigma_4 & \sigma_3 & \sigma_2 & \sigma_1  \\ \hline
{\hat m}_b:          & \sigma_{3'} & \sigma_{4'} & \sigma_{1'} & \sigma_{2'} & \sigma_3 & \sigma_4 & \sigma_1 & \sigma_2  \\ \hline
{\hat T}:     & \sigma_{1'} & \sigma_{2'} & \sigma_{3'} & \sigma_{4'} & \sigma_1 & \sigma_2 & \sigma_3 & \sigma_4  \\ \hline
{\hat T} {\hat I}:   & \sigma_{2'} & \sigma_{1'} & \sigma_{4'} & \sigma_{3'} & \sigma_2 & \sigma_1 & \sigma_4 & \sigma_3  \\ \hline
{\hat T} {\hat C}_{2b}: & \sigma_4 & \sigma_3 & \sigma_2 & \sigma_1 & \sigma_{4'} & \sigma_{3'} & \sigma_{2'} & \sigma_{1'}  \\ \hline
{\hat T} {\hat m}_b:   & \sigma_3 & \sigma_4 & \sigma_1 & \sigma_2 & \sigma_{3'} & \sigma_{4'} & \sigma_{1'} & \sigma_{2'}  \\ \hline
\end{array}$
\end{center}

\noindent Therefore, {\it eg}, $~\hat T \sigma_1^{NRXLD} = - \sigma_{1'}^{NRXLD}$ and similarly for all the other cases.

Using these relations we can then express $I_{LD}$ in term of one or two absorption site according to the symmetry group chosen. We consider the three cases of the groups: $P2/a+{\hat T}\{E|t_0\}P2/a$, $P2'/a$ and $P2/a'$ which are pertinent to the monoclinic phase.

In the case of the original group $P2/a+{\hat T}\{E|t_0\}P2/a$ (non ME), the global signal can be expressed in terms of only one independent center of absorption. Thus Eq. (\ref{intensity}) can be written as:

\begin{equation}
I_{LD} =  (1+{\hat {T}}{\hat {I}})(1+{\hat {T}}{\hat {m}}_b)(1+{\hat {T}}) \sigma_4^{LD}
\label{intensity1}
\end{equation}

In the case of a ME subgroups we use only the part of the table referring to the  operators of each subgroup so that

\begin{equation}
I_{LD}(P2'/a) =  (1+{\hat {T}}{\hat {I}})(1+{\hat {T}}{\hat {m}}_b)(\sigma_4^{LD}+ \sigma_{4'}^{LD})
\label{intensity2}
\end{equation}

\noindent and

\begin{equation}
I_{LD}(P2/a') =  (1+{\hat {T}}{\hat {I}})(1+{\hat {m}}_b)(\sigma_4^{LD}+ \sigma_{4'}^{LD})
\label{intensity3}
\end{equation}

\noindent in terms of two independent centers of absorption. In these formulas it is now intended that the group operators have already acted on the site but not on the physical observables. 

In all cases it is straightforward to derive that no circular dichroism can be 
detected for the system, as experimentally found in 
Ref. [\onlinecite{goulon}], since all the expressions are proportional to 
$(1+{\hat {T}}{\hat {I}})$ and the product of time-reversal and inversion 
symmetries is always -1 for circular dichroism 
($\hat {I}=1$ and $\hat {T}=-1$ in the E1-E1 channel and 
$\hat {I}=-1$ and $\hat {T}=1$ in the E1-E2 channel).
As expected, the three magnetic groups present different extinction rules for NRXLD. Since in this case the relevant operator is time-reversal and inversion odd, the signal is zero for the non ME group Eq. (\ref{intensity1}), due to the term $(1+{\hat {T}})$, and different from zero for the other two, since $\hat {T} \hat {I} = +1$.

On the contrary there is no case in which
the linear dichroism in the E1-E1 channel (NXLD) is forbidden by the crystal 
structure, because in this channel both the action of the time-reversal and the
action of the inversion operators give +1.  Thus, since the  magnetic symmetry around each vanadium ion is ${\hat {C}}_1$ (ie, no  symmetry at all) a non zero atomic NXLD is in principle expected. Such a dichroic signal is zero
only when each pair of vanadium ions contribute to the total intensity  
with an opposite sign. This latter occurrence is possible only 
for the particular direction at an angle $\pi/4$ with respect to the glide 
plane, in which case for each 
pair of vanadium ions related by this symmetry operation 
({\it eg} 1 and 3', 2 and 4', 3 and 1', 4 and 2', as inferred by the 
transformation table given above) 
$L^2_x\rightarrow L^2_y$ and viceversa, leading to the extinction of the 
total crystal signal as seen from Eqs. (\ref{intensity1}), (\ref{intensity2}) and (\ref{intensity3}). 
Note that any reduction of the symmetry to 
a group that does not contain the glide plane (either multiplied by the 
time-reversal or not) implies that the NXLD signal is never zero. 
Note also that the non-reciprocal signal, proportional to 
$(L^2_x - L^2_y) \Omega_z - h.c.$, at the particular 
angle $\pi$/4 vanishes if the magnetic group contains the glide-plane alone 
as a symmetry operation (like $P2/a'$) and may be different from zero with groups non 
containing it (like $P2'/a$).

We dwelt a bit on this analysis, since it highlights an unexpected feature 
in the nonreciprocal XLD spectra as presented in Ref. [\onlinecite{goulon}]. 
In fact, since one cannot disentangle the E1-E1 from the E1-E2 channel 
one would expect the superposition of two linear dichroic signals in any 
experimental spectrum, one of
which is time-reversal and inversion even (coming from  the E1-E1 transition) 
and a second, which is time-reversal and inversion odd (coming from the 
E1-E2 transition). As a consequence the residual signal obtained by taking 
the half sum of the two nonreciprocal XLD spectra, respectively with the 
external magnetic field oriented parallel/antiparallel to the hexagonal 
$c_H$ axis, ({\it ie} [XLD(H$^+$) + XLD(H$^-$)]/2), should give the normal 
(time-reversal and inversion even) linear dichroism signal, as correctly 
recognized by the authors in Ref. [\onlinecite{goulon}]. From Fig. 2 of the 
same reference one can infer that this residual signal is about ten times 
smaller than the nonreciprocal signal (which is itself between 0.0-1.0 \% of 
the main absorption), in practice nearly coinciding with the background noise, 
'due perhaps to the slightly distorted monoclinic structure', in their own 
words.

We show in the following section that a multiple scattering 
realistic simulation of the NXLD in the monoclinic lattice provides a 
signal which can be up to 4\% of the main absorption. The presence of such a signal is 
also corroborated by the sizable Templeton scattering at the 
(10$\overline{1}$)$_m$ forbidden reflection in the $\sigma-\sigma$ channel 
observed by Paolasini {\it et al.} \cite{paolasini2} in the monoclinic phase.
\cite{nota1} We performed a numerical simulation of the (10$\overline{1}$)$_m$ reflection (see next subsection) and found that the signal is almost entirely due to the E1-E1 channel.  Its existence implies a different value of the 
the $x$ and $y$ dipole matrix elements, which is at the basis of the NXLD. 
Notice that the Templeton scattering is zero in the corundum (paramagnetic) 
phase, as experimentally verified in the same work, \cite{paolasini2} due to 
the existence of the trigonal symmetry axis $C_3$, since then 
the $x$ and $y$ components of the electron scattering wavefunctions 
are degenerate in the real basis representation of the spherical harmonics. 
As a consequence mixed matrix elements of the kind $<\Psi_0|x|\Psi_n><\Psi_n|y|\Psi_0>$, which are at the origin of the Templeton effect in the $\sigma-\sigma$ channel at the (10$\overline{1}$)$_m$ reflection (see next subsection), vanish. 

The only possibility to have a situation like the one depicted in Ref. 
[\onlinecite{goulon}] is in the case that by mere accident the two 
orthogonal linear polarizations of the incoming photon beam in the geometrical setting of the experiment lay at an angle 
of $\pi/4$ relative to the monoclinic $b$ axis normal to the glide plane  
(whether or not multiplied by the time-reversal operator). 
At the same time to avoid the extinction of the NRXLD 
signal, the magnetoelectric group should be $P2'/a$, 
since in this case the operator ${\hat T \hat m_b}$ does not 
change the sign of $(L^2_x - L^2_y) \Omega_z - h.c.$.

 Under these assumptions the search for this particular geometrical setting would have directly determined the ground state symmetry of the AFI phase. However there is no trace in Ref. [\onlinecite{goulon}] of such investigation, so that their result cannot be unambiguously interpreted.

\subsection{Multiple Scattering simulations of NXLD}

The MS simulations of the normal linear dichroism in V$_2$O$_3$ were carried out in the framework of the multiple scattering theory within the muffin-tin approximation. We chose a cluster containing 135 atoms, {\it ie} 54 Vanadium and 81 oxygen atoms in the correct ratio 2 to 3, having a radius of 6.9 \AA, enough to get convergence with the cluster size. This latter is assured by convoluting the spectrum calculated with the real part of the Hedin-Lundqvist (HL) potential with a lorentian function having an energy dependent damping $\Gamma(E)$ derived from the universal mean free path curve by the relation $\lambda(E) = 1/k_e~E/\Gamma(E)$, where $E$ is the photoelectron kinetic energy (in Rydbergs) and $k_e = \sqrt{E}$ its wave vector. This procedure seems to give better results than calculating the spectrum directly with the complex HL potential, as recently found in other cases.\cite{benfatto} The actual calculations were performed independently with two different codes, the FDMNES program\cite{yves} which is non muffin-tin but incorporates a muffin-tin version and the CONTINUUM program,\cite{cnr} with similar results. As anticipated in the introduction, the independent particle approximation should be adequate to describe transitions to delocalized conduction states with $p$ symmetry around the photoaborber. Only strong quadrupole transitions to empty correlated $d$ states might not be well described, together with $p$ transitions to hybridized $p-d$ states in the same energy range. However we are only interested in the higher energy region where the nonreciprocal gyrotropic effect seems to be more pronounced. Moreover in the same energy range the muffin-tin approximation is reasonably accurate, due to the close-packed structure of V$_2$O$_3$ (despite the Vanadium voids) and the moderately high kinetic energy of the excited photoelectron. In order to eliminate any doubt on the confidence of the muffin-tin calculations, we also performed non-muffin-tin calculations on a substantially smaller cluster (eleven atoms) with the FDMNES program\cite{yves} and compared with the muffin-tin counterpart.
General shapes and, more important, amplitudes of the various signals discussed below did indeed compare very favorably. (Larger cluster calculations with no local symmetry, like in the present case, are probitively long in cpu time and require a lot of computer memory).

Even though the formalism of spherical tensors is very convenient to write down the response functions of x-ray spectroscopies, as illustrated in section III B, working in cartesian coordinates leads to immediately readable formulas.
In the geometry of Ref. [\onlinecite{goulon}] the direction of the incoming x-ray beam coincides with the trigonal axis of the high-temperature phase of V$_2$O$_3$. Thus, in the reference frame of Fig. 1, the two polarizations have components: $\epsilon_a \equiv  (\cos{\alpha},\sin{\alpha},0)$ and  $\epsilon_b \equiv (-\sin{\alpha},\cos{\alpha},0)$. 

The atomic linear dichroism in the E1-E1 channel is then given by:

\begin{eqnarray}
 \sigma_i^{NXLD} &\propto& \sum_n \delta(\omega_n) 
 \left( \langle \Psi^i_0|\vec{\epsilon}_a \cdot \vec{r}|\Psi_n \rangle 
\langle \Psi_n|\vec{\epsilon}_a \cdot \vec{r}|\Psi^i_0\rangle  \right. 
\nonumber  \\
             & - & \left. \langle \Psi^i_0|\vec{\epsilon}_b \cdot
\vec{r}|\Psi_n \rangle \langle \Psi_n|\vec{\epsilon}_b \cdot \vec{r}|\Psi^i_0
\rangle \right) \nonumber  \\
             & = & \sum_n \delta(\omega_n) 
\left\{ \cos 2\alpha (\langle x|x \rangle_i - \langle y|y \rangle_i )\right. 
\nonumber\\
 & + & \left. \sin 2\alpha Re[\langle x|y \rangle_i]\right\} 
\label{lindic2}
\end{eqnarray}

\noindent writing for brevity $\omega_n = \hbar\omega-(E_n-E_0)$, 
$\langle \Psi^i_0| x |\Psi_n \rangle \langle \Psi_n | x |\Psi^i_0 \rangle
\rightarrow \langle x|x \rangle_i$ and similarly for the other components.

When summing over all sites of the unit cell, only the part proportional to $\cos 2\alpha$ survives, since the mixed matrix elements $Re[\langle x|y \rangle_i]$ give opposite contributions for pairs related by the glide plane. Therefore the crystal absorption is given by:
\begin{equation}
\sigma_C^{NXLD}\propto \cos 2\alpha \sum_n \delta(\omega_n) (\langle x|x \rangle - \langle y|y\rangle )
\end{equation}

This is in contrast to what happens for Templeton scattering in the $\sigma-\sigma$ channel at the (10$\overline{1}$)$_m$ reflection in the same E1-E1 channel. In this case the incoming and outgoing polarization have components
$\epsilon_{\sigma} = (\cos \beta, \sin \beta, 0)$, so that the scattering amplitude for the atom at site $i$ reads:

\begin{eqnarray}
f_i(\omega) &=& (\hbar \omega)^2 \sum_n \frac{\omega_n+i\Gamma}{\omega_n^2 +
  \Gamma^2}\left(\cos^2\beta \langle x|x \rangle_i  
               +\sin^2\beta \langle y|y \rangle_i \right. \nonumber \\ 
            &+& \left. \sin2\beta Re[\langle x|y \rangle_i] \right)
\nonumber
\end{eqnarray}

Since the crystal structure factor at the (10$\overline{1}$)$_m$ reflection substracts the amplitudes coming from pairs of atoms related by the glide plane symmetry, the net crystal amplitude $F_C(\omega)$ turns out to be:

\begin{eqnarray}
F_C(\omega) \propto (\hbar \omega)^2 \sum_n \frac{\omega_n+i\Gamma}{\omega_n^2 +
  \Gamma^2} \sin 2\beta Re[\langle x|y \rangle] 
\end{eqnarray}

\noindent as anticipated above. Therefore in $C_3$ symmetry both linear dichroism and Templeton scattering at the (10$\overline{1}$)$_m$ reflection are zero, whereas if this latter does not vanish a normal linear dichroism is also expected.

\begin{figure}
\centerline{\epsfig{file=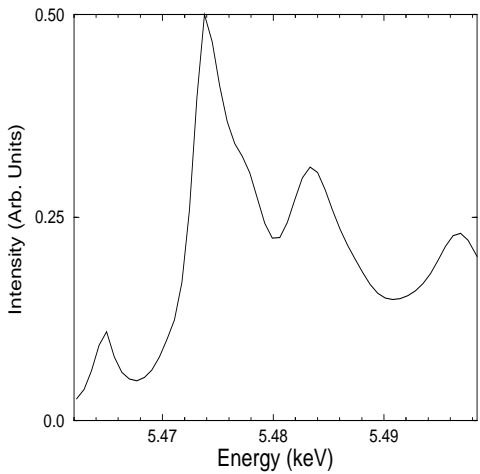,height=2.8in,width=2.8in}}
\vspace{5pt}
\caption
{Calculated Templeton intensity at the (10$\overline{1}$)$_m$ reflection. 
The ordinate shows the square of effective number of scattering electrons.}
\label{temple}
\end{figure}
  
However it is a bit hard to derive the size of the linear dichroism from the 
size of Templeton scattering, due to the extinction corrections in this latter 
spectroscopy. Our aim here is to show that the energy shape of the Templeton 
intensity is mainly of dipolar origin. To this purpose we present in Fig. \ref{temple} the E1-E1 contribution of our simulation for the $\sigma - \sigma$ channel at the  (10$\overline{1}$)$_m$ reflection. The agreement with the experimental spectrum of Ref. [\onlinecite{paolasini2}] is rather good, in the 
sense that we obtain peaks and valleys at the right energy position.
Only the intensities disagree with the experiment. However one should observe 
that fig.(2b) of Ref.  [\onlinecite{paolasini2}] represents the raw data not 
corrected for absorption. The amplitude of the E1-E2 contribution has been 
calculated to be ten times smaller than that in the E1-E1 channel, due to 
radial matrix elements effects and the very small weight of the $nd$ ($n>3$) 
scattering amplitudes in the conduction band of $p$ symmetry. Since in charge 
scattering the two contributions do not interfere, if one neglects magnetic 
order, the ratio of intensities in the two channels is 1 to one hundred.

\begin{figure}
\centerline{\epsfig{file=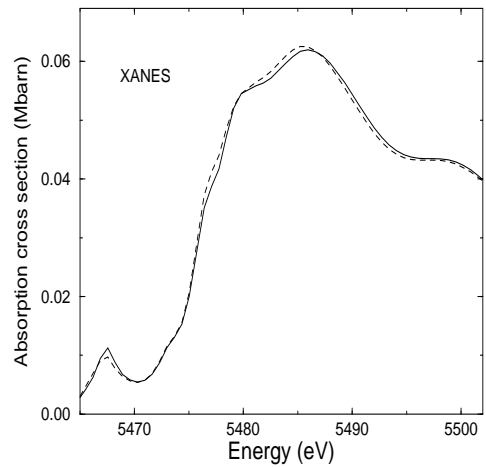,height=2.8in,width=2.8in}}
\vspace{5pt}
\caption
{Absorption signals (Mbarn) in V$_2$O$_3$ corresponding to polarizations within the hexagonal plane, parallel and perpendicular to b$_m$.}
\label{xanes}
\end{figure}

\begin{figure}
\centerline{\epsfig{file=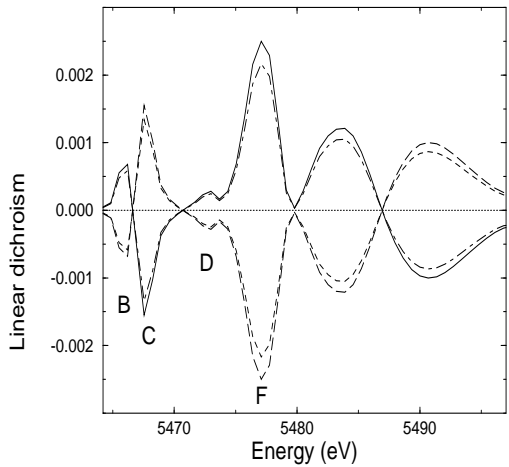,height=2.8in,width=2.8in}}
\vspace{5pt}
\caption
{Linear dichroism signals (Mbarn) as a function of the azimuthal angle $\alpha$ about the wave vector parallel to z. The four curves correspond to $\alpha = 0, 15, 75, 90^o$ degrees . The labeling of peaks corresponds to that of Ref. [15].}
\label{lindic}
\end{figure}

The main results of the numerical calculation for absorption and linear dichroism are shown in Figs. \ref{xanes} and \ref{lindic}. Figure \ref{xanes} shows the absorption signals in Mbarn with the incoming photon polarizations in the hexagonal plane and directed along b$_m$ and perpendicular to it. The general shape follows quite well the experimental counterpart.\cite{goulon} In Fig. \ref{lindic} instead we show on an expanded scale the variation of the dichroic signal (in Mbarn) as a function of the angle $\alpha$. The full line curve with negative maximum corresponds to $\alpha = 0^o$ ($<x|x> - <y|y>$), while the long-dashed line with positive maximum is for $\alpha = 90^o$ ($<y|y> - <x|x>$), in keeping with the $\cos2\alpha$ variation. The other two curves are for  $\alpha = 15^o$ and $\alpha = 75^o$ for later use. It is immedately evident that the dichroic signal vanish and then reverts its sign when the polarization angle crosses 45$^o$. Its shape is fairly similar to the non reciprocal experimental signal reported in Ref. [\onlinecite{goulon}] while the maximum changes from 0 to 4$\%$ of the main absorption and  is about 3$\%$ for $\alpha = 75 ^o$, to be compared with the experimental 1$\%$. In both figures we have not corrected for the 
unknown experimental resolution.

This finding is to be contrasted with an intensity more than 10 times smaller 
for the E1-E2 signal. This latter is in fact zero in the paramagnetic phase and picks 
up intensity only because of the magnetic ordering. Therefore on top of the 
factor one tenth in the matrix element, mentioned above when discussing 
Templeton scattering, one should add a further reduction factor originating 
from magnetic effects. We did not attempt to calculate the atomic signal as 
given by Eq. (\ref{lindice1e2}) for comparison, since the global signal 
depends on the crystal magnetic space group, as discussed above. Not knowing 
the mechanism that determines the lowering of the magnetic symmetry makes it 
impossible a meaningful estimate of the crystal magnetic effect. However this 
discussion should indicate that the non reciprocal gyrotropic effect in 
V$_2$O$_3$, if it exists, is more than one order of magnitude smaller than 
normal linear dichroism.

The relevance of these calculations to the argument we are discussing stems from the fact that there is a remote chance that the reversal of the dichroic signal with the magnetic field might be a geometrical artifact due to a lack of full control of the experimental conditions. In fact, since the procedure of magnetoelectric annealing is rather ineffective due to the non magnetoelectricity of V$_2$O$_3$ and the sizable conductivity of the sample in the paramagnateic insulating phase as discussed in Ref. \onlinecite{jansen}, it is not hard to conceive that the procedure performed in Ref. \onlinecite{goulon} could in reality involve two twins rotated by 120$^o$ degrees, so that the two curves refer to linear polarizations at 15$^o$ and 75$^o$ degrees with respect to the monoclinic b$_m$-axis. Therefore the control of the orientation of the crystal domain after annealing is essential in order to eliminate any ambiguity in the interpretation of the results.

Admittedly this is a quite remote possibility, however not more remote than the one depicted at the end of the previous subsection under the assumption that the observed effect was a truly nonreciprocal effect. Clearly the whole argument hinges on the existence of a sizable linear dichroic effect of E1-E1 character, which is anyway a prediction of the present work subject to experimental test. In fact a measurement of the  angular variation of the linear dichroism in the AFI phase without magnetoelectric annealing might settle the question.

\section{Conclusions}

In the preceding sections we have argued about the existence of a measurable normal ({\it ie} time-reversal and inversion even) linear dichroism in the AFI monoclinic phase of V$_2$O$_3$, both on the basis of realistic simulations in the framework of MS theory and experimental evidence coming from the observation of Templeton scattering at the (10$\overline{1}$)$_m$ forbidden reflection in the $\sigma-\sigma$ channel. \cite{paolasini2}  Therefore it comes to a surprise that no such a signal is present in the nonreciprocal linear dicroism experimental spectrum measured by Goulon {\it et al.}\cite{goulon} in the same sample.\cite{nota1} From the analysis of the crystal absorption cross section according to various ground state magnetic symmetry groups we have concluded that there exists only one group ($P2'/a$) and a particular geometrical setting capable of eliminating the NXLD signal and leaving in evidence the nonreciprocal dichroic effect.

Even accepting the fact that by a quite remote chance the authors in Ref. \onlinecite{goulon} had fallen on the correct setting, there still remain some paradoxical aspects with their result. The group $P2'/a$ is ME; however a recent measurement of the ME effect \cite{jansen} has given a negative result. Moreover this group does not seem to be among the possible candidates of the magnetic symmetry of the ground state in the AFI phase, as obtained by the minimization of an effective Hubbard Hamiltonian with degenerate bands.\cite{mila1,mila2,dimatteo} Finally, the reduction to the group  $P2'/a$ seems not compatible with the features of the Bragg-forbidden reflection (111)$_m$: the signal is present only at the quadrupole energies (5465 eV) and it is absent at the dipole energies (5470--5490 eV),\cite{paolasini1} while no extinction rules are present at the dipolar energies for the  $P2'/a$ group.\cite{yvesnext}

Faced to these paradoxes, we have tried to provide an alternative interpretation of the ``nonreciprocal'' results,\cite{goulon} invoking a geometrical origin for the reversal of the linear dichroism with the external magnetic field. However the question can only be settled by experimental tests in which one has the possibility to check the twin after each transition to the AFI phase, by means of an x-ray scattering equipment: this would give a full geometrical control of the system to be analyzed. Given the implications of the nonreciprocal effect, if confirmed, for the physics of V$_2$O$_3$, its unambiguous experimental determination is of the outmost importance.

\vspace{1cm}
%\acknoledgment

We would like to acknowledge interesting discussions with L. Paolasini on his measurements of the Templeton scattering in V$_2$O$_3$.

%\vspace{1cm}

\end{multicols}
\end{document}